\documentclass[11pt]{article}
\usepackage{amsmath,amsthm,amssymb,cite}
\usepackage{psfrag}
\usepackage{graphicx}
\newcommand{\nc}{\newcommand}
\nc{\mref}[1]{(\ref{#1})}
\nc{\vt}{v_{2\gL_0}}
\nc{\vo}{v_{\gL_0}}
\nc{\vot}{v_{\gL_1+\gL_0}}
\nc{\vw}{v_{\gL_1}}
\nc{\ppmm}{\genfrac{}{}{-10pt}{10pt}{++}{--}}
\nc{\wom}[5]{\Omega\left(\left.\begin{array}{ll}{#1}&{#2}\\{#3}&{#4}\end{array}\right|{#5}\right)}
\nc{\com}[5]{\chi\left(\left.\begin{array}{ll}{#1}&{#2}\\{#3}&{#4}\end{array}\right|{#5}\right)}
\nc{\we}[5]{W\left(\left.\begin{array}{ll}{#1}&{#2}\\{#3}&{#4}\end{array}\right|{#5}\right)}
\nc{\ce}[6]{C^{#6}\left(\left.\begin{array}{ll}{#1}&{#2}\\{#3}&{#4}\end{array}\right|{#5}\right)}
\nc{\lmat}[6]{\ell_{#6}\left(\left.\begin{array}{ll}{#1}&{#2}\\{#3}&{#4}\end{array}\right|{#5}\right)}
\nc{\lmats}[5]{L\left(\left.\begin{array}{ll}{#1}&{#2}\\{#3}&{#4}\end{array}\right|{#5}\right)}
\nc{\hmat}[6]{h_{#6}\left(\left.\begin{array}{ll}{#1}&{#2}\\{#3}&{#4}\end{array}\right|{#5}\right)}
\nc{\hmats}[5]{H\left(\left.\begin{array}{ll}{#1}&{#2}\\{#3}&{#4}\end{array}\right|{#5}\right)}
\nc{\web}[5]{\overline{W}\left(\left.\begin{array}{ll}{#1}&{#2}\\{#3}&{#4}\end{array}\right|{#5}\right)}
\nc{\wep}[5]{W'\left(\left.\begin{array}{ll}{#1}&{#2}\\{#3}&{#4}\end{array}\right|{#5}\right)}
\nc{\wes}[5]{W^*\left(\left.\begin{array}{ll}{#1}&{#2}\\{#3}&{#4}\end{array}\right|{#5}\right)}
\nc{\wess}[5]{W^{**}\left(\left.\begin{array}{ll}{#1}&{#2}\\{#3}&{#4}\end{array}\right|{#5}\right)}
\nc{\cet}[7]{C^{#6}_{#7}\left(\left.\begin{array}{ll}{#1}&{#2}\\{#3}&{#4}\end{array}\right|{#5}\right)}
\nc{\bcet}[7]{\bar{C}^{#6}_{#7}\left(\left.\begin{array}{ll}{#1}&{#2}\\{#3}&{#4}\end{array}\right|{#5}\right)}
\nc{\wet}[7]{W^{#6}_{#7}\left(\left.\begin{array}{ll}{#1}&{#2}\\{#3}&{#4}\end{array}\right|{#5}\right)}
\nc{\bwet}[7]{\overline{W}^{#6}_{#7}\left(\left.\begin{array}{ll}{#1}&{#2}\\{#3}&{#4}\end{array}\right|{#5}\right)}
\nc{\wec}[7]{\widetilde{W}^{#6}_{#7}\left(\left.\begin{array}{ll}{#1}&{#2}\\{#3}&{#4}\end{array}\right|{#5}\right)}
\nc{\wgen}[6]{W^{#6}\left(\left.\begin{array}{ll}{#1}&{#2}\\{#3}&{#4}\end{array}\right|{#5}\right)}
\nc{\wgenp}[6]{W^{*{#6}}\left(\left.\begin{array}{ll}{#1}&{#2}\\{#3}&{#4}\end{array}\right|{#5}\right)}
\nc{\wo}[5]{\Omega\left(\left.\begin{array}{ll}{#1}&{#2}\\{#3}&{#4}\end{array}\right|{#5}\right)}
\nc{\wsgen}[8]{{#8}^{#6}_{#7}\left(\left.\begin{array}{ll}{#1}&{#2}\\{#3}&{#4}\end{array}\right|{#5}\right)}
\nc{\qbinom}[2]{{\genfrac{[}{]}{0pt}{}{{#1}}{{#2}}}_{q}}
\nc{\hhg}[4]{\phi\left({{{#1}\,\,\,{#2}}\atop{{#3}}};
                     {#4}\right)}
\nc{\fullhhg}[5]{{_1}\phi_2\left({{{#1}\,\,\,{#2}}\atop{{#3}}};
                     {#4},{#5}\right)}
\nc{\bra}[1]{\langle #1 |}
\nc{\ket}[1]{| #1 \rangle}
\nc{\qp}[2]{({#1}\, ; \, {#2})_{\infty}}
\nc{\qpf}[1]{({#1}\, ; \, q^4)_{\infty}}
\nc{\pp}[1]{({#1}\, ; \, p)_{\infty}}
\nc{\qpp}[1]{({#1}\, ; \, p, q^4)_{\infty}}
\nc{\sect}{\section}
\nc{\ssect}{\subsection}
\nc{\sssect}{\subsubsection}
\nc{\ud}[1]{\underline{{#1}}}
\nc{\myra}[1]{\buildrel{{#1}}\over \longrightarrow}
\nc{\isomo}{\buildrel {\sim} \over \longrightarrow}
\nc{\Aff}{\operatorname{Aff}}
\nc{\ot}{\otimes}
\nc{\er}{\end{array}}
\nc{\bev}[1]{\begin{equation}\begin{array}{#1}}
\nc{\eeq}{\end{equation}}
\nc{\be}{\begin{eqnarray}}
\nc{\ee}{\end{eqnarray}}
\nc{\ben}{\begin{eqnarray*}}
\nc{\een}{\end{eqnarray*}}
\nc{\bec}{\begin{equation}\begin{array}{lll}}
\nc{\eec}{\end{array}\end{equation}}
\nc{\ed}{\end{document}}
\nc{\half}{\ensuremath{\frac{1}{2}}}
\nc{\Hom}{\operatorname{Hom}}
\nc{\End}{\operatorname{End}}
\nc{\vac}{|\textrm{vac}\rangle}
\nc{\tvac}{|\widetilde{\textrm{vac}}\rangle}
\nc{\dvac}{\langle\textrm{vac}}
\nc{\dtvac}{\langle\widetilde{\textrm{vac}}}
\nc{\id}{\mathbb{I}}
\nc{\ra}{\rightarrow}  
\nc{\lra}{\longrightarrow}
\nc{\uqp}{U^{\prime}_q (\widehat{sl}_2)}
\nc{\uqbp}{U_q (b_+)}
\nc{\uqbm}{U_q (b_-)}
\nc{\ub}{U^{\prime}_q (b_+)}
\nc{\vsl}{V(\sigma(\lambda))}
\nc{\vl}{V(\lambda)}  
\nc{\bu}{\bullet}
\nc{\an}{{\ell}}
\nc{\slth}{\widehat{\mathfrak{sl}}_2\hskip 1pt}
\newcommand{\uq}{U_q\bigl(\slth\bigr)}
\nc{\ws}{\;\;}
\nc{\qu}{{1\ov 4}}
\nc{\hif}{\hb{ if }}
\nc{\hev}{\hb{ is even }}
\nc{\hod}{\hb{ is odd }}
\nc{\Tr}{{\rm Tr}}
\nc{\ad}{{\rm Ad}}
\nc{\hb}{\hbox}
\nc{\nn}{\nonumber} 
\nc{\curlra}{\buildrel{\sim}\over\longrightarrow}
\nc{\epp}{\varepsilon^{\prime}} 
\nc{\ol}{\overline}
\nc{\pl}{\prod\limits} 
\nc{\sli}{\sum\limits} 
\nc{\nin}{\noindent}

\nc{\ga}{\alpha}
\nc{\gb}{\beta}
\nc{\gd}{\delta}
\nc{\gep}{\varepsilon}
\nc{\gz}{\zeta}
\nc{\gt}{\theta}
\nc{\gk}{\kappa}
\nc{\gl}{\lambda}
\nc{\gp}{\phi}
\nc{\gs}{\sigma}
\nc{\go}{\omega}
\nc{\gn}{\nu}
\nc{\gr}{\rho}

\nc{\gou}{\underline{\go}}
\nc{\un}{\underline{n}}
\nc{\um}{\underline{m}}
\nc{\uw}{\underline{w}}

\nc{\s}{\sigma}
\nc{\ep}{\varepsilon}
\nc{\z}{\zeta}
\nc{\g}{\gamma}
\nc{\zi}{\zeta^{-1}}

\nc{\gG}{\Gamma}
\nc{\gD}{\Delta}
\nc{\gT}{\Theta}
\nc{\gL}{\Lambda}
\nc{\gO}{\Omega}
\nc{\gP}{\Phi}

\nc{\cL}{\mathcal{L}}
\nc{\cF}{\mathcal{F}}
\nc{\cP}{\mathcal{P}}
\nc{\cS}{\mathcal{S}}
\nc{\cN}{\mathcal{N}}
\nc{\cD}{\mathcal{D}}
\nc{\cH}{\mathcal{H}}
\nc{\cO}{\mathcal{O}}
\nc{\cT}{\mathcal{T}}
\nc{\cQ}{\mathcal{Q}}
\nc{\cW}{\mathcal{W}}
\nc{\cR}{\mathcal{R}}

\nc{\C}{\mathbb{C}}
\nc{\Q}{\mathbb{Q}}
\nc{\R}{\mathbb{R}}
\nc{\Z}{\mathbb{Z}}
\nc{\N}{\mathbb{N}}

\nc{\fg}{\mathfrak{g}}

\nc{\barx}{\bar{x}}
\nc{\bi}{\bar{i}}
\nc{\bj}{\bar{j}}
\nc{\bgr}{\bar{\rho}}
\nc{\bA}{\bar{\alpha}}
\nc{\bB}{\bar{\beta}}
\nc{\bC}{\bar{\gamma}}
\nc{\by}{\bar{y}}
\nc{\brv}{\overline{V}}
\nc{\brp}{\overline{P}}

\nc{\tf}{\tilde{f}}
\nc{\te}{\tilde{e}}
\nc{\ts}{\tilde{s}}
\nc{\tgP}{\widetilde{\Phi}}
\nc{\tgPs}{\tilde{\Psi}}
\nc{\tgn}{\tilde{\nu}}
\nc{\tgl}{\tilde{\lambda}}
\nc{\tge}{\tilde{\eta}}
\nc{\txi}{\tilde{\xi}}
\nc{\tep}{\tilde{\epsilon}}

\nc{\cB}{\check{b}}
\nc{\cOm}{\check{\Omega}}

\nc{\goto}{\mapsto}
\nc{\embed}{\hookrightarrow}
\nc{\rien}{\emptyset}
\nc{\lb}[1]{\label{#1}}
\nc{\Nt}{\frac{N}{2}}
\nc{\vn}{\hspace*{-33truemm}}
\nc{\vm}{\hspace*{-0truemm}}
\nc{\ti}{t^{-1}}
\nc{\vb}{v^{(1)}}
\nc{\vbn}{v^{(n)}}
\nc{\ur}{\underline{r}}
\nc{\us}{\underline{s}}
\nc{\up}{\underline{p}}
\nc{\bp}{\bar{p}}
\nc{\bpi}{\bar{p}^{(i)}}
\nc{\bpip}{\bar{p}^{(i+1)}}
\nc{\vz}{V^{(1)}_z}
\nc{\vzn}{V^{(n)}_z}
\nc{\vzo}{V^{(1)}_1}
\nc{\piz}{\pi_z^{(1)}}
\nc{\pizn}{\pi_z^{(n)}}
\nc{\pis}{\pi_{(z,\us)}}
\nc{\bW}{\overline{W}}
\nc{\bQ}{\overline{Q}}
\nc{\tQ}{\widetilde{Q}}
\nc{\bT}{\overline{T}}
\nc{\note}[1]{\vspace*{-5mm}\marginpar[left]{\scriptsize\bf{#1}}}
\nc{\eqdef}{:=}
\nc{\lu}{^{(\lambda)}}
\nc{\vone}{v_{\Lambda_1}}
\nc{\vzero}{v_{\Lambda_0}}

\def\bea{\begin{eqnarray}}
\def\eea{\end{eqnarray}}

\def\ep{\varepsilon}

\newtheorem{theorem}{Theorem}[section]

\newtheorem{example}[theorem]{Example}
\def\by{{\mbox{\boldmath $y$}}}

\def\bpm{\begin{pmatrix}}
\def\epm{\end{pmatrix}}
\def\bdet{\left|\begin{array}}
\def\edet{\end{array}\right|}

\nc{\dx}[1]{\frac{d{#1}}{dx}}
\nc{\ddx}[1]{\frac{d^2{#1}}{dx^2}}
\nc{\dt}[1]{\frac{d{#1}}{dt}}
\nc{\ddt}[1]{\frac{d^2{#1}}{dt^2}}

\newcounter{quest}
\newlength{\questoffset}
\nc{\pd}{\partial}
\nc{\nl}{\newline}
\nc{\vp}{\vspace*{3mm}}
\nc{\soln}{\noindent {\it Solution}:\,}
\nc{\pro}{\noindent {\it Proof}\,:\,}
\nc{\bex}{\begin{example}}
\nc{\eex}{\end{example}}

\nc{\mitem}{\noindent}

\newcommand{\ena}{\end{eqnarray}}

\def\cip(#1){(#1;p,q^4)_\infty}
\def\z{\zeta}

\def\Lpm#1{\mathrel{\mathop{\kern0pt L^\pm}\limits^#1}}
\def\L#1{\mathrel{\mathop{\kern0pt L}\limits^#1}}
\def\Lm#1{\mathrel{\mathop{\kern0pt L^-}\limits^#1}}
\def\Lp#1{\mathrel{\mathop{\kern0pt L^+}\limits^#1}}

\begin{document}
\bibliographystyle{unsrt}
\begin{flushright}
\end{flushright}
\begin{center}
{\LARGE \bf An Algebraic Setting for Defects in the \\ XXZ and Sine-Gordon Models }\\[10mm]
{\large \bf 
Robert Weston}\\[3mm]
{\it Department of Mathematics, Heriot-Watt University,\\
Edinburgh EH14 4AS, UK.\\
{\tt  R.A.Weston@ma.hw.ac.uk}}\\[3mm]
June 2010\\[10mm]
\end{center}
\begin{abstract}
\noindent 
We construct defects in the XXZ and sine-Gordon models by making use of the representation theory of $\uq$. The representations involved are generalisations of the infinite-dimensional, $q$-oscillator representations used in the construction of $Q$-operators. We present new results for intertwiners of these representations, and use them to consider both quantum spin-chain Hamiltonians with defects and quantum defects in the sine-Gordon model. We connect specialisations our results with the work of Corrigan and Zambon on type $I$ and type $I\!I$ defects, and present sine-Gordon soliton/defect and candidate defect/defect scattering matrices.
\end{abstract}
\nopagebreak

\section{Introduction}
Integrable quantum field theory and solvable lattice models started out as separate fields, the one dealing with high-energy and the other with condensed matter physics. The two fields merged as they were subsumed within the larger theory of quantum inverse scattering. The level of understanding of
 such  models has deepened as the number of examples studied and applications 
considered has grown enormously. 
Heightened interest in this area in the last few years can be explained by a few factors: the number of new, exact results that have been found \cite{JM,KMT05}, the appearance of quantum spin-chains in string theory \cite{MR1976005}, and
the use of exact results in the description of experimental, quasi-one-dimensional systems \cite{Caux08}. 

Two aspects of these systems that have received attention in the recent renaissance
 are $Q$-operators and defects. In this paper, we demonstrate that there is a connection between these two types of object. The reason for this connection, expressed in the technical language of the quantum inverse scattering method, is that they are both constructed in terms of the monodromy matrix with an infinite-dimensional auxiliary space. The goals of this paper are to explain this statement to non-experts and to exploit the technology developed in the study of $Q$-operators  to produce a catalogue of results applicable to  defects. 

Baxter introduced $Q$-operators as a technical tool in his solution of the 8-vertex model\cite{Bax72a}; they allowed him to derive Bethe equations for Hamiltonian eigenvalues in the absence of a Bethe ansatz for the eigenvectors. This use of $Q$-operators became redundant when Baxter introduced another clever trick \cite{Bax73aI,Bax73aII,Bax73aIII} that enabled him to obtain eigenvectors\footnote{Baxter invented and used a mapping of the 8-vertex model to the SOS model.}. The $Q$-operator then fell into obscurity before resurfacing into the mainstream with the work of Bazhanov, Lukyanov and Zamolodchikov (BLZ). BLZ made use of a $Q$-operator in their construction of the transfer matrix in quantum field theory \cite{MR1384140,BLZb,Bazhanov:1998dq,MR1832065}. Their insight was that the $Q$-operator was a more fundamental object that the transfer matrix. In fact, the transfer matrix could be constructed  in terms of their $Q$. BLZ constructed the $Q$-operator by using a monodromy matrix for $\uq$ that involved an auxiliary space that was an infinite-dimensional, $q$-oscillator representation of a Borel subalgebra. This approach was  developed and generalised by 
various authors \cite{Rossi:2002ed,MR2184024,MR1880098,MR2426018}.

The interest in both conformal field theories and massive integrable models has followed a similar route. Bulk models were studied first \cite{MR757857}, followed 
by models with boundaries \cite{MR858661}, and finally models with defects \cite{MR1279145,MR1674671}. Models with boundaries and defects are of interest for various reasons: physical systems have them; they affect the field content; they increase the richness and complexity of the models; and they are an essential aspect of the branes/string theory paradigm. However, while there is a large body of literature on integrable models with boundaries, much less work has been done on integrable defects.

 One motivation for the current work was to develop an algebraic understanding of the work  of  Bowcock, Corrigan and Zambon (BCZ) on type $I$ defects \cite{MR2087101,MR2045593,MR2165823,MR2326786}, and the more recent work of Corrigan and Zambon (CZ) on fused, or type $I\!I$, defects \cite{CoZa09,CZ10}.\footnote{We use the type $I/II$ nomenclature of \cite{CZ10}.}
BCZ start from a classical Lagrangian density for the sine-Gordon model with
 an integrable defect inserted at the spatial origin. This defect produces interesting field equations
 for the sine-Gordon field on the left and right of the defect. Away from the defect these two fields obey independent conventional sine-Gordon equations. At the defect, the two fields are related by a classical Backlund transformation. The authors consider the classical scattering of solitons with this defect. As solitons pass through the defect their topological charge can stay the same, or they can flip from soliton to anti-soliton and deposit topological charge on the defect. The defect thus carries an odd or even integer topological charge. It also carries a rapidity-like parameter. Making use of this classical scattering data, the authors go on to conjecture a corresponding quantum soliton/defect scattering matrix. They also take steps towards producing a defect/defect scattering matrix \cite{MR2165823}. That such an objects exist
 might seem odd, but the defects considered in \cite{MR2165823} can move independently and can thus scatter. 

The classical, type $I$ defects of BCZ have subsequently been studied and generalised by several authors who have made use of the Lax pair description and solved a Riccati-like equation to obtain conserved charges\cite{MR2398297,MR2469290}. The quantum defects have also been considered within a more algebraic framework in \cite{Baj06,Baj08,MR2398297}.
In the recent work of CZ \cite{CoZa09,CZ10}, the type $I$ defects of BCZ have been generalised to a parameter-dependent type $I\!I$ defect.

The approach of BCZ to finding the quantum soliton/defect scattering matrix is to solve a Yang-Baxter equation of the form $S TT =TTS$, where $S$ is the soliton $S$ matrix and $T$ is the desired soliton/defect scattering matrix.\footnote{The $T$ matrix for type $I$ sine-Gordon defects was first obtained in this way in \cite{MR1674671}.} In this paper, we compute $T$ in a more general setting by solving a simpler, linear equation. We do this in the time-honoured, quantum inverse scattering way by converting the problem into one of representation theory. In this approach, $S$ becomes the $R$-matrix which is the intertwiner for the spin-1/2 representation  $V_{\z}$ of the quantum affine algebra $\uq$: 
\ben 
R(\z_1/\z_2): V_{\z_1}\ot  V_{\z_2} \ra  V_{\z_1}\ot  V_{\z_2}.\een
The $T$ matrix becomes a special case of a more general intertwiner $\cL$ of an infinite-dimensional representation $W^{(\ur)}_{\z_1}$ and the above $V_{\z_2}$:
\ben \cL(\z_1/\z_2): W^{(\ur)}_{\z_1} \ot V_{\z_2} \ra W^{(\ur)}_{\z_1} \ot V_{\z_2} .\een This new Borel subalgebra representation $W^{(\ur)}_{\z}$ is parametrised by a rapidity-like parameter $\z$ and by a complex 
vector $\ur=(r_0,r_1,r_2)$, and is a generalisation of the $q$-oscillator representations that have been used in the construction of the $Q$-matrix. It is a slightly modified version of the representation introduced in \cite{Rossi:2002ed}. The $Q$-matrix would be given as a regularised trace over a product of such $\cL$ operators. In the application to defects, it is $\cL$ itself that is of interest.
Finding $\cL$ amounts to solving the linear intertwining condition $\cL \Delta (x) = \Delta'(x) \cL$, where $\Delta(x)$ is the Borel subalgebra
coproduct and $\Delta'(x)$ is defined in Section 2. We find that the defect/soliton scattering matrix for both type $I$ and type $I\!I$ defects can be expressed in terms of different specialisations of our $\cL$ operator. The connection between the $\cL$ operator of \cite{Rossi:2002ed} and type $I$ defects was considered previously in \cite{Baj06}.

In Section 2 of this paper, we  consider the required representation theory of $\uq$ and construct the various intertwiners that we will associate with defects. 
We  go on to consider the application to defects in the 6-vertex and XXZ models in Section 3. 
This connection is of course natural since $R(\z)$ specifies the Boltzmann weights of the 6-vertex model.  In Section 4, we make the connection with the work of CZ. Specialisations of $\ur$ \, for our defect reproduce the defect/soliton scattering results of CZ for both type $I$ and type $I\!I$ defects. 
We also consider defect/defect scattering. Finally, we summarise our work in Section 5.

\section{$\uq$ Representation Theory and Defects}
In this section, we develop the representation theory that we shall use to describe defects and their interactions. The starting point is the quantum affine algebra $\uqp$, in terms of which both the XXZ $R$-matrix and sine-Gordon S-matrix may be written. We will not reproduce too many details of this algebra, which can be found in many other places \cite{ChPr91}.
However, it simplifies the subsequent discussion of the Borel subalgebra representations if we give the relations: the algebra $\uqp$ is an associative algebra over the complex numbers generated by the elements $e_i,f_i,t_i^{\pm 1}$ with relations\footnote{The prime on $\uqp$ indicates that we are not including a
derivation in the definition.} \\[-5mm]
\bea
&&[e_i,f_j]=\delta_{i,j} \frac{t_i-\ti_i}{q-q^{-1}},\\
&& t_i e_i \ti_i = q^2 e_i, \quad t_i e_j \ti_i = q^{-2} e_j \quad (i\neq
j),\lb{reln1}\\
&&t_i f_i \ti_i = q^{-2} f_i, \quad t_i f_j \ti_i = q^{2} f_j \quad (i\neq
j),\\
&&e_i e_j^3 -[3] e_j e_i e_j^2 + [3] e_j^2 e_i e_j - e_j^3 e_i=0 \quad (i\neq j),\lb{Serre}\\
&&f_i f_j^3 -[3] f_j f_i f_j^2 + [3] f_j^2 f_i f_j - f_j^3 f_i=0 \quad
(i\neq j).\eea
We use the coproduct $\gD: \uqp \ra \uqp \ot \uqp$ given by
\bea
\gD(e_i) = e_i \ot 1 + t_i \ot e_i,\quad \gD(f_i) = f_i \ot t_i^{-1} + 1
\ot f_i,\quad \gD(t_i)=t_i\ot t_i.\eea
The Borel subalgebra $\uqbp$ that we consider is the one generated by the elements $e_i,t_i^{\pm 1}$. The only relevant relations from the above are therefore \mref{reln1} and the Serre relation \mref{Serre}. In the rest of this section we present representations and the associated intertwiners for both these algebras.
\subsection{The Generalised Oscillator Algebra}
We define a generalised oscillator algebra that we shall use in order to construct $\uqbp$ representations. Let $r_1,r_2$ be complex numbers. Then, we define 
the generalised oscillator algebra, $\cO^{(r_1,r_2)}$, to be the associative algebra generated by
$a$, $a^*$, $q^{\pm D}$ with relations
\ben q^D a^* q^{-D}= q a^*,\ws q^D a q^{-D}= q^{-1} a,\ws
a a^* = (r_1+q^{-2D}) (r_2 + q^{2D}),\ws a^* a = (r_1+q^{2-2D}) (r_2 + q^{2D-2}).\een
 Note that we recover the more conventional $q$-oscillator algebras \cite{Ku91} when either $r_1=0$ or $r_2=0$, in which cases we have the $q$-oscillator relations
\ben a^* a - q^2 a a^* = (1-q^2),\ws \hbox{or} \quad a a^* - q^{2} a^* a= (1-q^2)\quad
\hb{respectively}.\een

We consider the $\cO^{(r_1,r_2)}$ module $W^{(r_1,r_2)}=\oplus_{j\in \Z} \C \ket{j}$ defined by 
\ben
a\ket{j}= \ket{j-1}, \quad a^*\ket{j}=(r_1+q^{-2j}) (r_2+q^{2j}) \ket{j+1}, \quad
q^{\pm D}\ket{j}=q^{\pm j}\ket{j}.\een

\subsection{Borel Subalgebra Representations}
Let $\ur$ denote the vector $(r_0,r_1,r_2)\in \C^3$. Let   $W^{(\ur)}_\z$ be the infinite-dimensional $\uqbp$ module, spanned by
$\ket{j}\ot \z^n \in W^{(r_1,r_2)}\ot \C[[\z,\z^{-1}]]$, with $\uqbp$ action 
\ben && e_0 (\ket{j}\ot \z^n) =\frac{1}{q-q^{-1}} a^* \ket{j}\ot \z^{n+1},\quad
 e_1 (\ket{j}\ot \z)=\frac{1}{q-q^{-1}} a \ket{j}\ot \z^{n+1}, \\ &&t_1 (\ket{j}\ot \z^{n}) = r_0 \,q^{-2D} \ket{j}\ot \z^n,\quad t_0 (\ket{j}\ot \z^n) = r_0^{-1} q^{2D} \ket{j}\ot \z^n.\een
The  module $W^{(\ur)}_\z$ is a convenient reparametrisation of the $\uqbp$ module first introduced, and shown to be the most general solution of the relations \mref{reln1} and \mref{Serre}, in the paper \cite{Rossi:2002ed}. 
\subsubsection{Special cases}\lb{scases}
In the special case when either $r_1=-q^{-2n}$ or $r_2=-q^{2n}$ ($n\in \Z$), we have $a^*\ket{n}=0$, and the $\uqbp$ module is modified to $(\oplus_{j\leq n} \C \ket{j})\ot \C[[\z,\z^{-1}]]$.
We obtain further truncation of $W^{(\ur)}_\z$ in the cases $\ur=(q^n,-q^{-2n},-q^{-2})$ or $\ur=(q^n,-q^2,-q^{2n})$ to the finite module
$\oplus_{j=0}^{n} \C \ket{j})\ot \C[[\z,\z^{-1}]]$. Furthermore, in these cases there are 
respective $\uqbp$ isomorphisms to $V^{(n)}_{\z q^{-(n+1)/2}}$ and  $V^{(n)}_{\z q^{(n+1)/2}}$. Here, $V^{(n)}_\z$ denotes
the spin-$n/2$ principal evaluation module which generalises the spin-$1/2$ module $V_\z=V^{(1)}_\z$ defined in Section 2.3.
 In the root of unity case $q^{2N}=1$,
the module  $W^{(\ur)}_\z$ may also be truncated to ($\oplus_{j=0}^{N-1} \C \ket{j})\ot \C[[\z,\z^{-1}]].$

\subsection{Evaluation Representations}
We will make use of the $\uqp$ principal evaluation module $V_{\z}=(\C v_+ \oplus \C v_-)\ot \C[[\z,\z^{-1}]]$ defined by\\[-9mm]
\ben 
&& e_0 (v_+ \ot \z^n) =  (v_- \ot \z^{n+1}),\quad  e_1 (v_- \ot \z^n) =  (v_+ \ot \z^{n+1}),\quad 
e_0 (v_- \ot \z^n) =0,\quad  e_1 (v_+ \ot \z^n)=0\\
&& f_0 (v_- \ot \z^n) =  (v_+ \ot \z^{n-1}),\quad  f_1 (v_+ \ot \z^n) =  (v_- \ot \z^{n-1}),
\quad f_0 (v_+ \ot \z^n) = 0,\quad f_1 (v_- \ot \z^n) =0, \\
&& t_0 (v_{\pm} \ot \z^n) =  q^{\mp 1} (v_+ \ot \z^n),\quad  t_1 (v_\pm \ot \z^n) =  q^{\pm 1}(v_\pm \ot \z^n).\een
The R-matrix $R(\z_1/\z_2): V_{\z_1}\ot  V_{\z_2} \ra  V_{\z_1}\ot  V_{\z_2}$ that obeys
 $R(\z)\circ \Delta(x) = \Delta'(x) \circ R(\z)$ for all $x\in \uqp$ is given by\footnote{If $\Delta(x)=\sli_{i} a_i\ot b_i$, then $\Delta'(x)=\sli_i b_i\ot a_i$.}
\bea
 R(\z)= \frac{1}{\kappa(\z)}\bar{R}(\z),\quad  
 \bar{R}(\z)=\bpm 1 & 0 & 0 & 0 \\
 0&\frac{(1-\z^2)q}{1-q^2\z^2} & \frac{(1-q^2)\z}{1-q^2\z^2}&0\\[2mm]
 0&\frac{(1-q^2)\z}{1-q^2\z^2} & \frac{(1-\z^2)q}{1-q^2\z^2}&0\\
 0 & 0 & 0 & 1 
\epm.
\lb{erm1}\eea
We do not give the explicit expression for $\kappa(\z)$ in this paper, but this, and the crossing and unitarity properties of $R(\z)$ that $\kappa(\z)$ ensures, can be found in Appendix A of \cite{JM}. 

\begin{figure}[htbp]
\centering
\includegraphics[width=2cm]{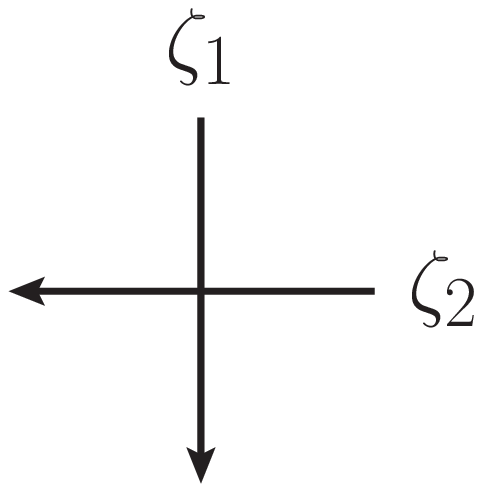}
\caption{The Operator $R(\z_1/\z_2)$}
\label{frm1}
\end{figure}

In order to understand the properties of $R(\z)$ and other intertwiners it is very useful to work with pictures; we represent $R(\z)$ as shown in Figure \ref{frm1}.
The arrows indicate that the operator acts from the North-East to the South-West. 

\subsection{The L-operator}
\begin{figure}[htbp]
\centering
\includegraphics[width=2cm]{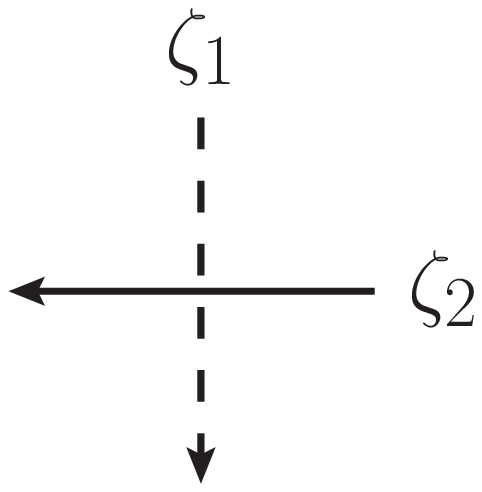}
\caption{The Operator $\cL^{}(\z_1/\z_2)$ }
\label{lop}
\end{figure}

We now define a new object: a $\uqbp$ intertwiner $\cL^{(\ur)}(\z_1/\z_2): W^{(\ur)}_{\z_1} \ot V_{\z_2} 
\ra W^{(\ur)}_{\z_1} \ot V_{\z_2} $ that obeys \bea
\cL(\z)\circ \Delta(x) = \Delta'(x) \circ \cL(\z)\lb{edeltal}\eea for all $x\in \uqbp$. Solving this linear equation, we find (with a particular choice of normalisation) 
the expression
\bea
\cL^{(\ur)}(\z,q)=     \bpm (1+r_2 \z^2 q^{2-2D})&  -\z a^*  \\ -\z a  & (1+r_1 \z^2 q^{2D})
 \epm \bpm q^{D} &0\\0& r_0 \, q^{-D}\epm 
\label{elm1}.\eea
In writing \mref{elm1} in the form shown, and in finding the intertwiner $\cR$ in Section \ref{goaint}, we have been led by the approach of the appendices of the papers \cite{MR2472021,BJMST07} that deal with the $q$-oscillator 
cases.\footnote{An $\cL$ operator which is similar to \mref{elm1} appears without derivation in the paper \cite{MR1688868}.}
Except when we specifically require it, we will suppress both the $\ur$ superscript and $q$ argument of $\cL^{(\ur)}(\z,q)$, and write it as
$\cL(\z)$.
We represent the infinite-dimensional module $W^{(\ur)}(\z_1/\z_2)$ by a dashed line, and depict $\cL(\z_1/\z_2)$ by Figure \ref{lop}.

\subsubsection{Properties of $\cL(\z)$}
\begin{figure}[htbp]
\centering
\includegraphics[width=7cm]{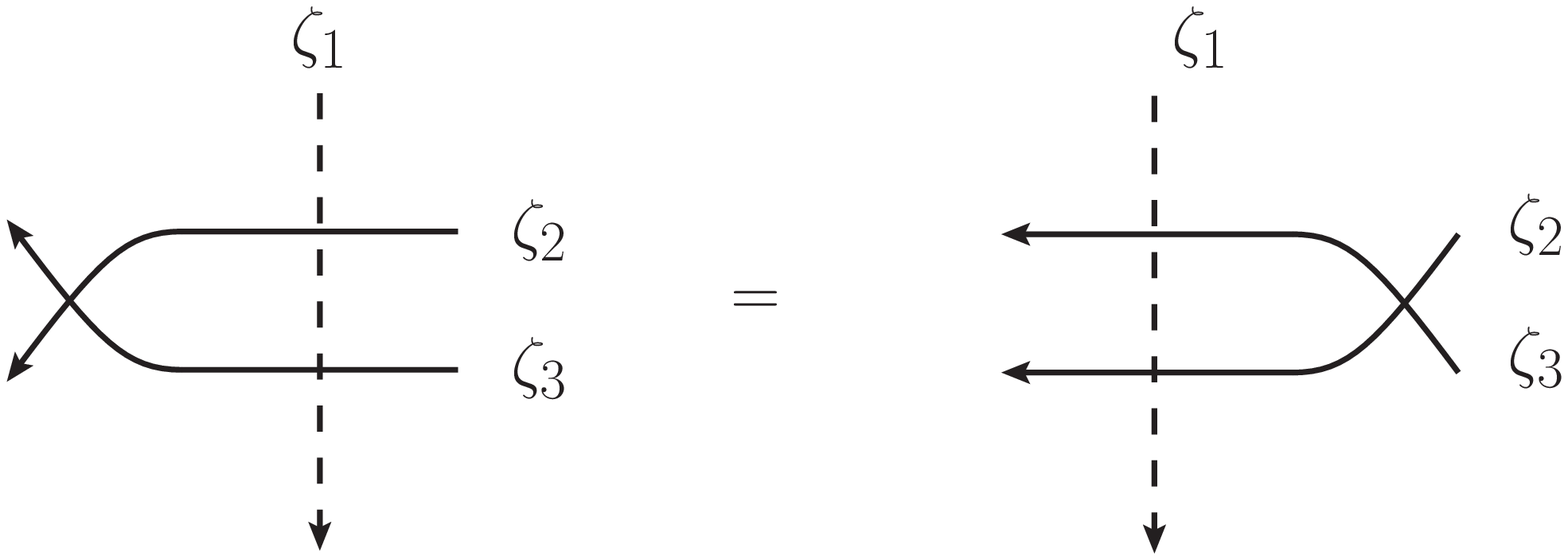}
\caption{The Yang-Baxter relation on $W^{(\ur)}(\z_1) \ot V_{\z_2}\ot V_{\z_3}$}
\label{ybf1}
\end{figure}

We have, by construction, the Yang-Baxter relation
 \bea R_{2,3}(\z_2/\z_3) \cL_{1,3}(\z_1/\z_3)  \cL_{1,2}(\z_1/\z_2) = \cL_{1,2}(\z_1/\z_2) \cL_{1,3}(\z_1/\z_3)   R_{2,3}(\z_2/\z_3)\lb{eyb1},\eea
where both sides act on the space $W^{(\ur)}(\z_1) \ot V_{\z_2}\ot V_{\z_3}$. The relation is indicated by Figure \ref{ybf1}. 

We find that the inverse operator is given by
\ben 
\cL^{-1}(\z)=\frac{1}{(1-\z^2)(1-\z^2 r_1 r_2)} \bpm q^{-D} &0\\0& r_0^{-1}\,q^{D}\epm 
\bpm (1+r_1 \z^2 q^{2D-2})&  \z a^* \\ \z a  & (1+r_2 \z^2 q^{-2D}) \epm.
\een
The operator $\cL(\z)$ also obeys  the crossing relation
\ben
\cL^{-1}(-\z q)=\frac{1}{r_0(1-\z^2 q^2)(1-\z^2 q^2 r_1 r_2)} \big(\sigma^x \cL(\z) \sigma^x)^{t_2}
\een
where $\gs^x$ is the Pauli matrix, and $t_2$ indicates transpose with respect to the two dimensional space.  
 
In the introduction, we mentioned the connection of our work with $Q$-operators.  The 
substance of this connection is that we can identify a $Q$ operator
as a suitably regularised trace of $\cL(\z)$ over the space $W^{(\ur)}_\z$ \cite{Rossi:2002ed,MR2184024}. The resulting operator obeys the $TQ=Q+Q$ Baxter relation as a result of  the fusion properties that
of $W^{(\ur)}(\z)$ and $V_\z$ that we will now discuss. It can be shown, following the method of \cite{Rossi:2002ed}, that we have the short exact sequence
\ben
0\lra W^{(qr_0,q^{-2} r_1,r_2)}_{\z q} \myra{\iota} W^{(r_0,r_1,r_2)}_\z \ot V_{\z}\myra{\pi}  W^{(q^{-1} r_0,q^{2} r_1,r_2)}_{\z q^{-1}}\lra 0.\een
This constitutes a concise statement of the fact that $W^{(qr_0,q^{-2} r_1,r_2)}_{\z q}$ is a submodule of $W^{(r_0,r_1,r_2)}_\z \ot V_{\z}$ and that there is an isomorphism
$ W^{(r_0,r_1,r_2)}_\z \ot V_{\z}/W^{(qr_0,q^{-2} r_1,r_2)}_{\z q} \simeq W^{(q^{-1} r_0,q^{2} r_1,r_2)}_{\z q^{-1}}$. 
The embedding $\iota$ and projection $\pi$ are given explicitly by 
\ben  \iota: W^{(qr_0,q^{-2} r_1,r_2)}_{\z q} &\ra& W^{(r_0,r_1,r_2)}_\z \ot V_{\z}\\
 \ket{j} &\goto& A_j:=r_0(q^{j-1}r_1+q^{1-j})\ket{j}\ot v_+ + q^j \ket{j-1} \ot v_- , \quad\hb{and},\\[3mm]
\pi: W^{(r_0,r_1,r_2)}_\z \ot V_{\z}&\ra& W^{(q^{-1}r_0,q^{2} r_1,r_2)}_{\z q^{-1}}  \\
  \ket{j}\ot v_+   &\goto  &  q^j \ket{j-1},\\
A_j&\goto  & 0.
\een
A consequence of this short exact sequence is that the $\cL$ operator for $ W^{(q^{\pm 1} r_0,q^{\mp 2} r_1,r_2)}_{\z q^{\pm 1}}$ is related to that of $W^{(r_0,r_1,r_2)}_\z$ via a fusion relation. With our choice of normalisation, we find
\ben
(\iota \ot 1)\, \cL^{(qr_0 ,q^{-2}r_1,r_2)}(\z q)&=&  \frac{(1-q^2 \z^2)}{(1-\z^2)} \cL^{(r_0,r_1,r_2)}_{1,3}(\z) \bar{R}_{2,3}(\z)
(\iota \ot 1),\\
 \cL^{(q^{-1}r_0 ,q^{2}r_1,r_2)}(\z q^{-1}) (\pi \ot 1)&=& q^{-1} (\pi \ot 1) \cL^{(r_0,r_1,r_2)}_{1,3}(\z) \bar{R}_{2,3}(\z),
\een
where the first relation acts on  $W^{(qr_0,q^{-2} r_1,r_2)}_{\z_1 q} \ot V_{\z_2}$, the second acts on 
$W^{(r_0,r_1,r_2)}_{\z_1} \ot V_{\z_1}\ot V_{\z_2}$, and $\z=\z_1/\z_2$.
We include these relations because of the potential application in the study of soliton/defect fusion.

\subsection{Intertwiners of Generalised Oscillator Representation}
\begin{figure}[htbp]
\centering
\includegraphics[width=2cm]{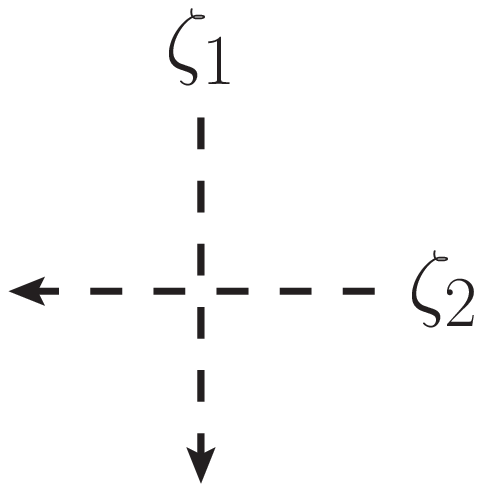}
\caption{The Intertwiner $\cR(\z_1/\z_2)$ }
\label{drm}
\end{figure}

\begin{figure}[htbp]
\centering
\includegraphics[width=7cm]{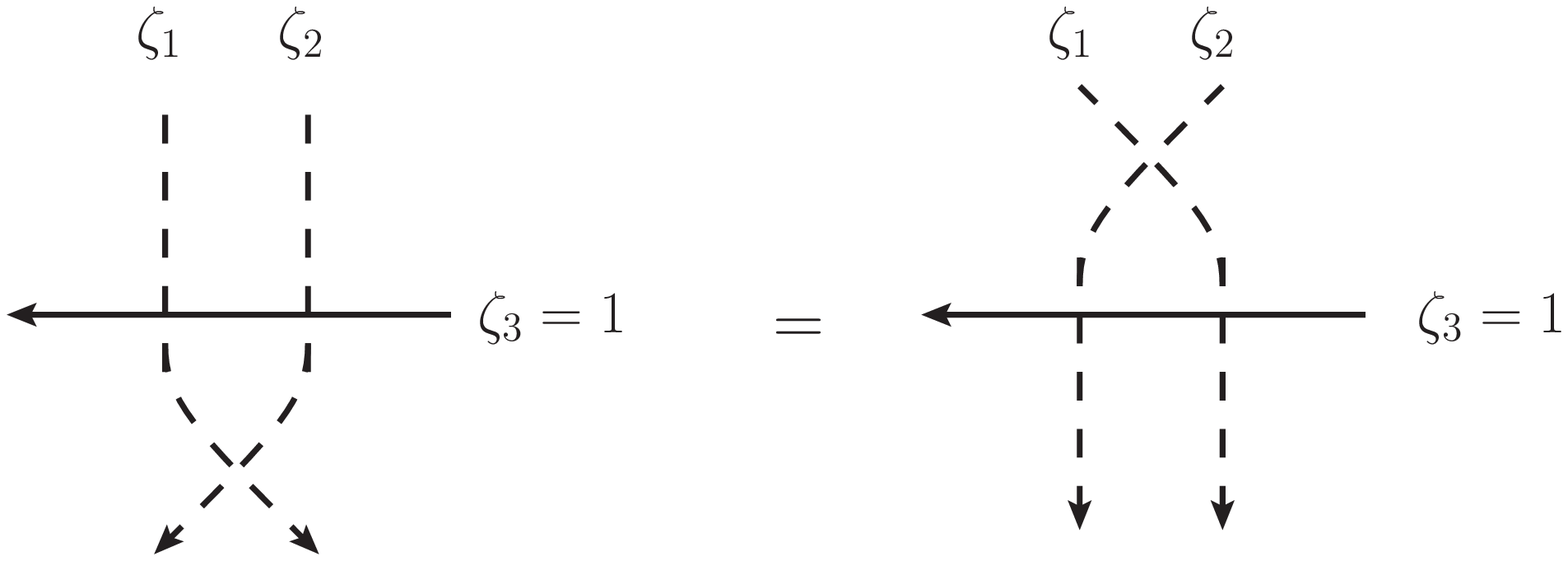}
\caption{The Representation of Equation \mref{ddreln}}
\label{ybf2}
\end{figure}

\lb{goaint}
In this subsection, we consider intertwiners of the generalised oscillator representations of $\uqbp$. That is, we look for an intertwiner $\cR^{(\ur)}(\z_1/\z_2): W^{(\ur)}_{\z_1}\ot W^{(\ur)}_{\z_2}
 \ra W^{(\ur)}_{\z_1}\ot  W^{(\ur)}_{\z_2}$ that obeys the condition 
 \bea \cR^{(\ur)}_{1,2}(\z_1/\z_2) \cL^{(\ur)}_1(\z_1)  \cL^{(\ur)}_2(\z_2) = \cL^{(\ur)}_2(\z_2) \cL^{(\ur)}_1(\z_1)   \cR^{(\ur)}_{1,2}(\z_1/\z_2)\lb{ddreln},\eea
where  $1$ and $2$ denote the first and second components, and  $\cL^{(\ur)}_1(\z_1)  \cL^{(\ur)}_2(\z_2)$ indicates $2\times 2$
matrix multiplication of the $\cL$-matrices.  We have for example that 
\ben
\big(\cL^{(\ur)}_1(\z_1)  \cL^{(\ur)}_2(\z_2)\big)_{+,+}= (q^{D_1} + r_2 \z_1^2 q^{2-D_1})(q^{D_2} + r_2 \z_2^2 q^{2-D_2}) +\z_1 \z_2 r_0 a^*_1 q^{-D_1}a_2 q^{D_2}.\een 
Pictorially, we have $\cR(\z_1/\z_2)$ and the relation \mref{ddreln} given by Figures \ref{drm} and \ref{ybf2}.

We do not have a solution of \mref{ddreln}  for $\cR^{(\ur)}(\z)$ for generic $\ur$, but we do have  formal solutions when either $r_1=0$ or $r_2=0$. 
We obtain these by extending the method of \cite{MR2472021}, which deals with this problem in the  $r_1=-1$, $r_2=0$ case.
We make use of the the exchange matrix $P(\ket{j} \ot \ket{k}) = (\ket{k} \ot \ket{j})$.
Following the approach of Appendix A of \cite{MR2472021}, we write the intertwining matrices in the form
\bea \cR^{(r_0,r_1,0)}(\z)= P\, h(\z,r_0 \, u) \, \z^{D_1+D_2}, \quad 
 R^{(r_0,0,r_2)}(\z)= P\, h(\z,r_0^{-1} u') \, \z^{-(D_1+D_2)}\lb{ersoln}\eea
where $u=a_1^* q^{-2D_1} a_2$, $u'=a_1 q^{2D_1} a^*_2$,  and $h(\z,v)$ 
is a formal series in $v$. In both the $r_1=0, r_2\neq 0$ and $r_1\neq 0, r_2=0$ cases, Equation \mref{ddreln} then reduces to the following single condition for the formal series:
\bea h(\z,v)(1+\z v)&=& h(\z,q^2 v) (1+\z^{-1} v).\lb{fps}\eea
If we assume that $h(\z,0)=1$, then Equation \mref{fps} has the unique solution
\ben
 h(\z,v)&=&\sli_{n\geq 0} (- q^{-1} v)^n \prod\limits_{m=1}^n \frac{(\z^{-1} q^{m-1} -\z q^{1-m})}{q^m-q^{-m}}.
 \een
Using the q-binomial theorem, this can be rewritten for $|v\zeta|<1$, and $|q|<1$ as\footnote{The infinite product is defined as $(a;b)_\infty=\prod\limits_{n=0}^\infty (1-a\,b^n)$.} 
\ben h(\z,v)=\frac{(-v\z^{-1};q^2)_\infty}{(-v\z;q^2)_\infty}.\een

We have solved \mref{ddreln} to obtain $\cR^{(\ur)}(\z)$ for the most general $\ur$ cases that we can. The expense of this generality is that the solutions \mref{ersoln} are formal series in $u$ or $u'$. However, in the special cases $r_1=-q^{-2n}$ or $r_2=-q^{2n}$ discussed in Section \ref{scases}, the operators $u$ and $u'$ are nilpotent and the series are well defined. In the more restrictive case $r_1=0$, $r_2=0$, to be considered in connection with the work of BCZ in our Section 4, we have two well-defined, independent solutions given by \mref{ersoln}.

\section{Defects in the 6-Vertex and XXZ Models}
\begin{figure}[htbp]
\centering
\includegraphics[width=2cm]{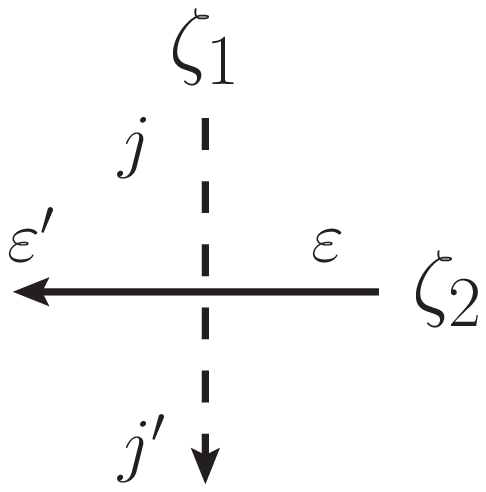}
\caption{The Boltzmann Weight $\cL(\z_1/\z_2)^{j,\ep}_{j',\ep'}$}
\label{bw}
\end{figure}

The components of the $R$-matrix \mref{erm1} specify the vertex weights of the 6-vertex model. We can use the $\cL(\z)$ 
operator to define a new Boltzmann weight, that is we can define components of $\cL(\z)$ by
\ben \cL(\z) (\ket{j}\ot v_{\ep})= \sli_{j',\ep'}\cL(\z)^{j,\ep}_{j',\ep'} (\ket{j'}\ot v_{\ep'}),\een
where $\ep,\ep'\in\{+1,-1\}$, and $j\in \Z$ (at least in the generic $\ur$ case). Note that the only non-zero contribution to the above sum comes from
 $j'=j+(\ep'-\ep)/2$.
These components may then be associated with the Boltzmann weight for the edge configuration of Figure \ref{bw}.
We can use this weight to introduce a defect line into the 6-vertex model as shown in Figure \ref{pf}. The simplest scenario is to assume an $N\times N$ 
finite lattice with periodic boundary conditions. 
\begin{figure}[htbp]
\centering
\includegraphics[width=4cm]{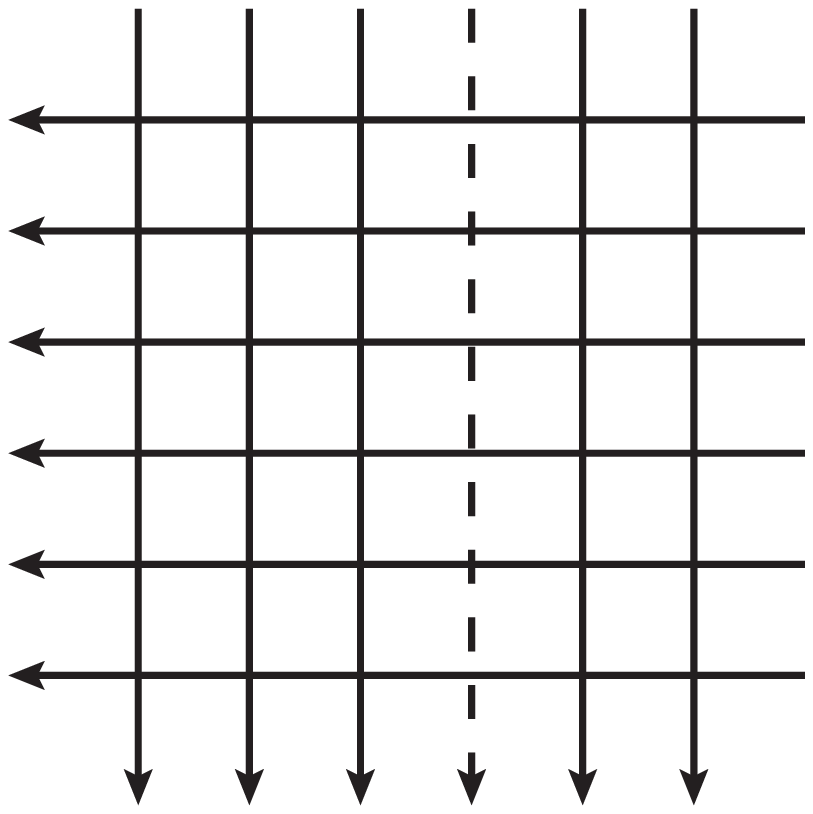}
\caption{The 6-vertex model with a defect line}
\lb{pf}
\end{figure}

\begin{figure}[htbp]
\centering
\includegraphics[width=5cm]{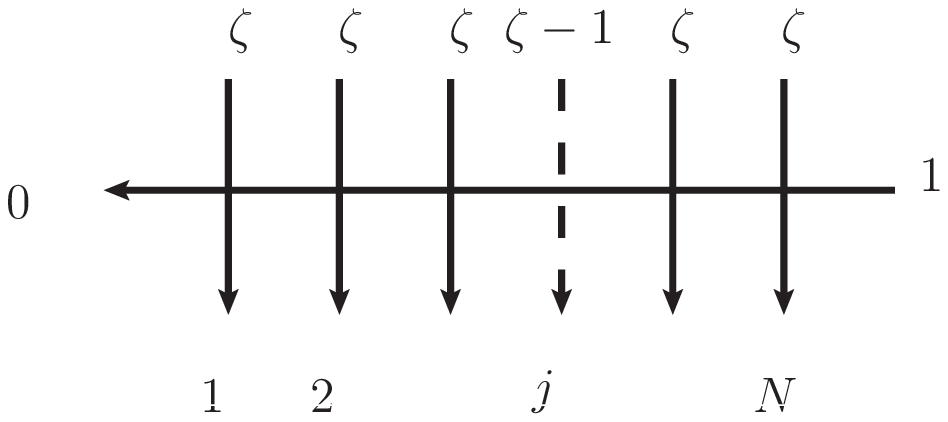}
\caption{The transfer matrix $T(\z)$ for the XXZ model with a defect}
\label{tm}
\end{figure}

We can consider an anisotropic 6-vertex model with a defect with different $\z$ `spectral'  parameters on different vertical and horizontal lines. In fact, in order to produce a simple quantum spin-chain Hamiltonian, it is convenient to consider the case when all horizontal lines have spectral parameter 1, all vertical lines with the exception of the defect line have spectral parameter $\z$, and the defect line itself has a spectral parameter $\z-1$. The horizontal transfer
matrix associated with this choice is
\ben T(\z)=\Tr_{\C^2}\big(R_{1,0}(\z) R_{2,0}(\z) \cdots R_{j-1,0}(\z) \cL_{j,0}^{(\ur)}(\z-1) R_{j+1,0}(\z)\cdots R_{N-1,0}(\z) R_{N,0}(\z)\big).\een
This operator acts on the space $\cH= \C^2 \ot \C^2 \ot \cdots \ot \C^2 \ot W^{(r_1,r_2)}\ot \C^2\cdots \ot \C^2\ot \C^2 $ and the trace is over the
two dimensional horizontal auxiliary space labelled by `0'. This transfer matrix is represented in Figure \ref{tm}. The solvability/integrability property $[T(\z),T(\z')]=0$ follows from the Yang-Baxter equation for $R$ and from Equation \mref{eyb1}.

\begin{figure}[htbp]
\centering
\includegraphics[width=2cm]{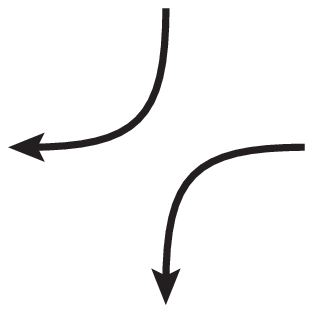}
\caption{Representation of $R(1)=P$}
\label{fp}
\end{figure}

\begin{figure}[htbp]
\centering
\includegraphics[width=2cm]{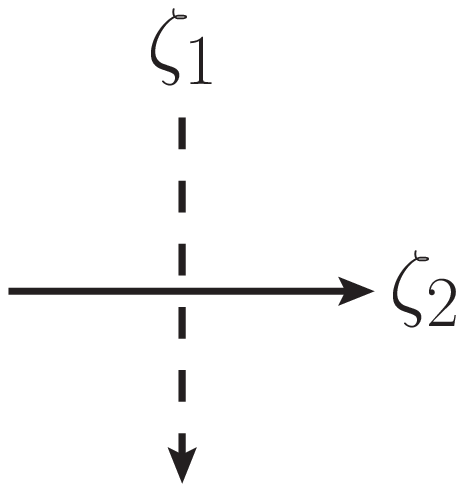}
\caption{Representation of $\cL^{-1}(\z_1/\z_2)$}
\label{linv}
\end{figure}

\begin{figure}[htbp]
\centering
\includegraphics[width=3cm]{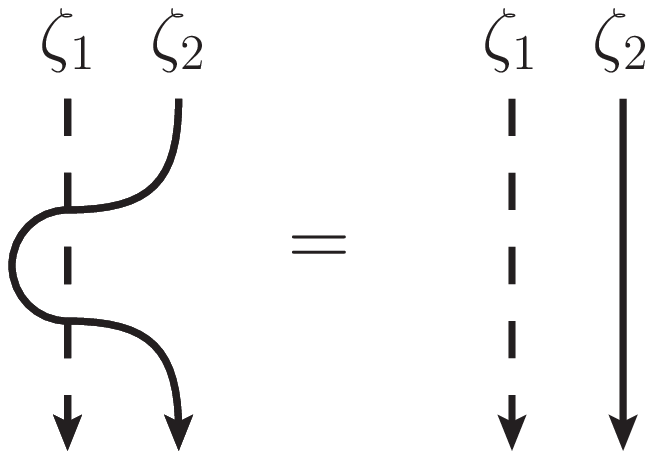}
\caption{Representation of $\cL^{-1}(\z_1/\z_2) \cL(\z_1/\z_2)=1$}
\label{l-linv}
\end{figure}

The XXZ Hamiltonian with a defect is given in terms of the logarithmic derivative of $T(\z)$ at $\z=1$. One of the goals of this paper is to communicate the mysteries of the quantum inverse scattering method to a non-expert readership, and so rather than just writing down the Hamiltonian, we will give a derivation, emphasising the simplifying role of pictures. The starting point is to note that  $R(1)=P$, the permutation operator\footnote{The normalisation function $\kappa(\z)$  has the property $\kappa(1)=1$.}. This is represented graphically by Figure \ref{fp}.
We also note that $\cL^{-1}(\z_1/\z_2)$ can be represented by Figure \ref{linv} with the relation $\cL^{-1}(\z_1/\z_2)\cL(\z_1/\z_2)=1$ represented by 
Figure \ref{l-linv}.


\begin{figure}[htbp]
\centering
\includegraphics[width=3cm]{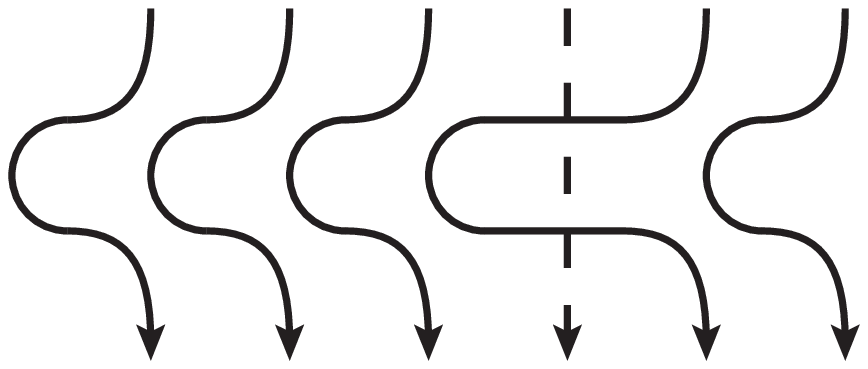}
\caption{Representation of $T(\z^{-1}) T(\z)=1$}
\label{t-tinv}
\end{figure}

\begin{figure}[htbp]
\centering
\includegraphics[width=4cm]{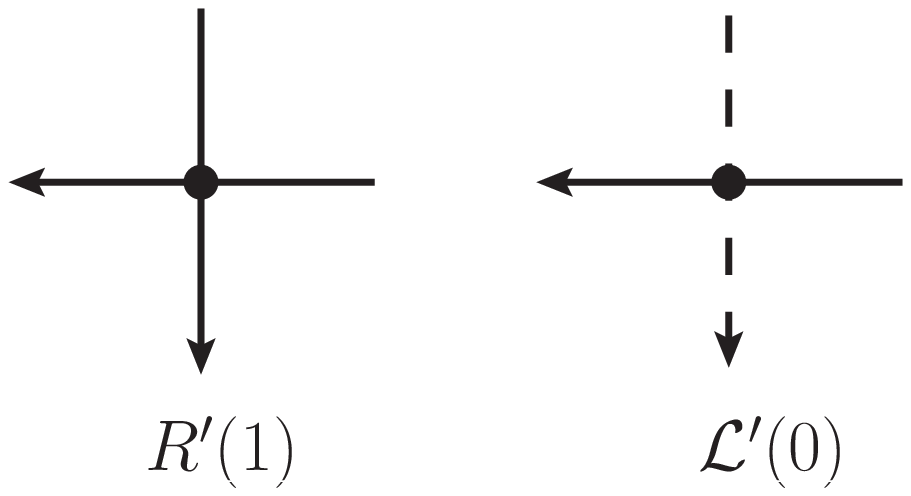}
\caption{Representation of $R'(\z=1)$ and $\cL'(\z=0)$}
\label{diffvert}
\end{figure}

\begin{figure}[htbp]
\centering
\includegraphics[width=8cm]{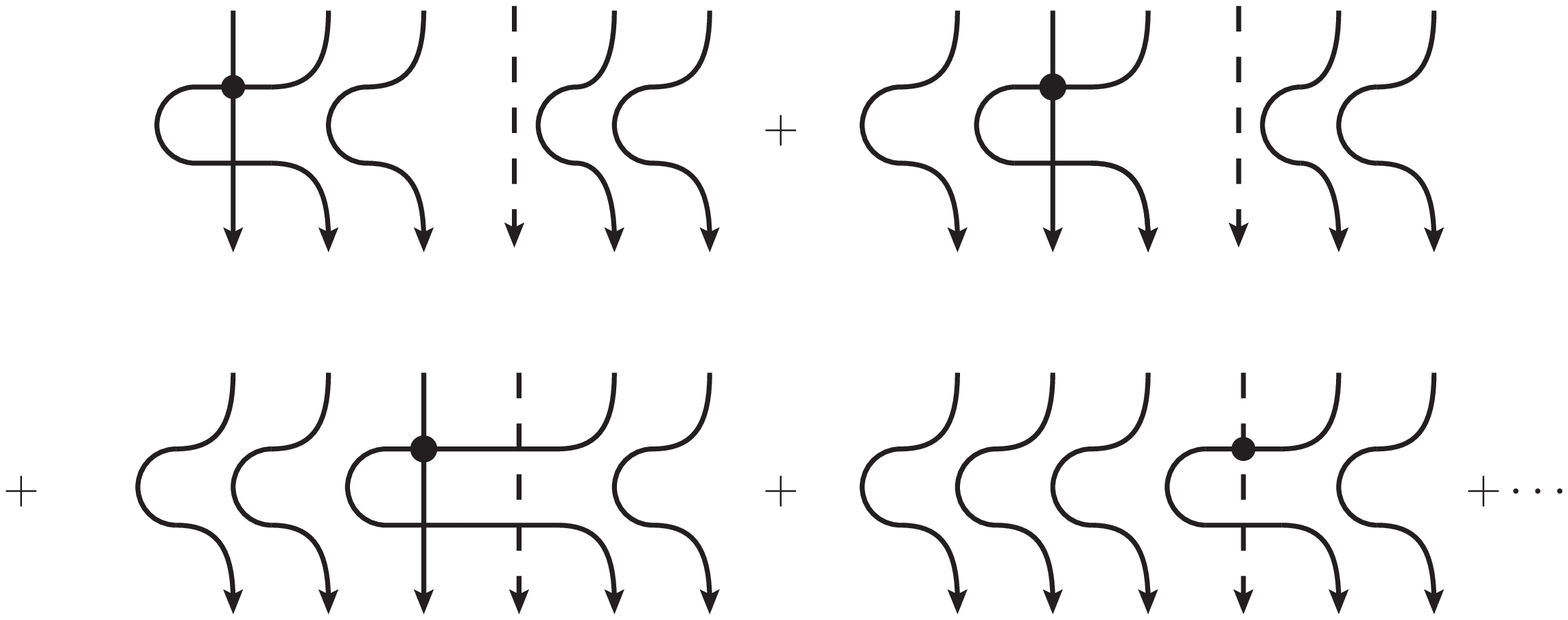}
\caption{Representation of $\frac{d \ln(T(\z))}{d\z}\big|_{\z=1}$}
\label{hamilt}
\end{figure}

Hence, the relation $T(1)^{-1} T(1)= 1$ can be represented by Figure \ref{t-tinv}.
The Hamiltonian however is given in terms of $\frac{d \ln(T(\z))}{d\z}\big|_{\z=1}=T^{-1}(\z) T'(\z)\big|_{\z=1}$. 
Let us represent $R'(\z=1)$ and $\cL'(\z=0)$ by the vertices with black bullets as shown in Figure \ref{diffvert}.
Then finally we arrived at the pictorial representation of $\frac{d \ln(T(\z))}{d\z}\big|_{\z=1}$ shown in Figure \ref{hamilt}.

We can then essentially read off the Hamiltonian from Figure \ref{hamilt} after noting the explicit form of $R'(1)$ and $\cL'(0)$. We have
\ben 
&& \frac{d \ln(R(\z))}{d\z}\Big|_{\z=1} =-\kappa'(1) +\frac{q\Delta}{1-q^2}+\frac{2q}{1-q^2}h,\quad  \\[2mm]
\hb{where}&& h:=
-\frac{1}{2} (\gs^x\ot \gs^x +  \gs^y\ot \gs^y+ \Delta \, \gs^z\ot \gs^z) =
\bpm 
-\frac{\Delta}{2}\\[-1mm]
&\frac{\Delta}{2}&-1\\
&-1&\frac{\Delta}{2}\\[-1mm]
&&&-\frac{\Delta}{2}\epm,\\
\hb{and} &&\frac{d \ln(\cL^{(\ur)}(\z))}{d\z}\big|_{\z=0} = - \bpm 0 & r_0 a^*\, q^{-1-2D} \\ r_o^{-1} a\, q^{2D-1}& 0 \epm.\een
If we define $H=\frac{1-q^2}{2q} \frac{d\ln(T(\z)}{d\z}\big|_{\z=1}$, we have the integrable Hamiltonian for the XXZ model with a defect:
\ben H=const+ \hspace*{-3mm}\sli_{i\neq j-1,j} \hspace*{-3mm} h_{i,i+1} + 
\bpm q^{-D} &0\\0&  \hspace*{-2mm}r_0^{-1}q^{D}\epm_{j,j+1}\hspace*{-8mm}
h_{j-1,j+1}\bpm q^{D} &0\\0&  \hspace*{-2mm}r_0 q^{-D}\epm_{j,j+1} \hspace*{-6mm} +
\frac{q-q^{-1}}{2q} \bpm  0 & \hspace*{-4mm} r_0 a^*\, q^{-2D} \\ r_0^{-1} a\, q^{2D}& 0 \epm_{j,j+1} \hspace*{-8mm}. \een
Note, that the Hamiltonian for the XXZ model without a defect is simply $\sli_{i} h_{i,i+1}$.

\section{Defects in the Sine-Gordon Model}
BCZ have considered type $I$ defects in the sine-Gordon and other affine-Toda models in a series of papers \cite{MR2087101,MR2045593,MR2165823,MR2326786}. CZ have generalised parts of this work to a parameter-dependent type $I\!I$ defect in 
\cite{CoZa09,CZ10}. The starting point in both the original and generalised cases is a classical sine-Gordon Lagrangian with a delta function term localised at the spatial origin. The authors match solitonic solutions on either side of the defect, and interpret the results as classical, reflectionless, soliton/defect  scattering. They go on to consider quantum scattering, 
and follow the S-matrix programme of solving quantum Yang-Baxter-like equations and finding a solution that is consistent with the classical data. 
Schematically, if $S$ represents the known soliton/soliton scattering matrix and $T$ the soliton/defect transmission matrix, then BCZ solve a Yang-Baxter equation of the form $STT=TTS$ to find $T$. This is a difficult quadratic relation to solve, even for the relatively simple scattering process considered in \cite{MR2165823}. In Section 3 of this paper, we have presented an algebraic scheme. In this section, we will  identify $S$ with our $R$ and the soliton/defect scattering matrix for both type $I$ and type $I\!I$ defects with our $\cL$. Rather than solving \mref{eyb1} directly, we have had the luxury of solving the much simpler relation \mref{edeltal}. 

Let us first summarise the results of CZ concerning the sine-Gordon model \cite{CZ10}.\footnote{We adopt the conventions of the paper \cite{CZ10} which differ slightly from the earlier works of these authors.} Their soliton S-matrix is 
\cite{MR2165823}\ben
 S(\gt,q)=\rho_s(\gt) \bpm qx-x^{-1} q^{-1} \\
& q-q^{-1}& x-x^{-1}\\
&  x-x^{-1}&q-q^{-1}\\
&&&qx^{-1}-x^{-1} q^{-1} \epm, \quad x=e^{\gamma \gt},\quad q=-e^{-i \pi \gamma},
\een
where $\gt$ is the rapidity and $\gamma$ is related to the conventional sine-Gordon coupling constant by
$\gamma={\frac{4\pi -\gb^2}{\gb^2}}$. Type $I$ defects are characterised by a continuous parameter $\eta$ and by an integer topological charge (denoted by lower case Greek letters $\ga$ or $\gb$). Type $I\!I$ defects are characterised by two complex parameters $b_\pm$ and and by the same integer topological charge. The conjectured soliton/defect transmission matrices, $T_{I}(\gt,\eta)$ and $T_{II}(\gt,b_+,b_-)$, are given by Equations (3.4) and (3.15) of \cite{CZ10}. 
The identification of the results of \cite{CZ10} and the objects of Section 3 of this paper is relatively straightforward. First of all, we identify
\bea \frac{1}{(qx-q^{-1}x^{-1})\rho_s(\gt) }S(\gt,q)= P\bar{R}(x,q). \lb{sident}\eea
Then, we find 
\bea \frac{1}{\rho_I(\gt,\eta)}T_I(\gt,\eta) &=& \nu^{-\half}\, U_I(\nu)\,\, \cL^{(r_0=\nu,r_1=0,r_2=0)} (i e^{\gamma(\gt-\eta)},q)\, \,U_I^{-1}(\nu).\lb{tident}\\
\frac{1}{\rho_{I\!I}}T_{I\!I}(\gt,b_+,b_-) &=& U_{I\!I}(b+,b-)\,\, \cL^{(r_0=1,r_1=\frac{\overline{b}_- q^2}{\overline{b}_+},r_2=\frac{{b}_-}{q^2 b_+} )} ( i e^{\gamma \gt} |b_+|,q) \,\, U_{I\!I}^{-1}(b+,b-),\lb{tiident}\eea
where the gauge transformations are given by
\ben U_I(\nu)=\bpm 1&0\\0& - \nu^\half q^{-D-\half}\epm, \quad U_{I\!I}(b_+,b_-) =\bpm 1&0\\0&
 \frac{i}{|b_+|}(b_+ + b_-q^{-2D-2})\epm.\een
Note that we also identify $\ga=-2j$.\footnote{We have also taken into account transposed matrix conventions: in this paper we use $ {\cal O} \ket{a}=\sli_b {\cal O}^a_b \ket{b}$, whereas in \cite{CZ10}, they use $ {\cal O} \ket{a}=\sli_b {\cal O}^b_a \ket{b}$.} These gauge transformation do not affect the $R\cL \cL=\cL \cL R$ relations of Equation \mref{eyb1}, i.e., we have the relations with the original $R$ and with the gauge transformed $\cL$.
Hence, the $STT=TTS$ soliton-defect transmission-matrix  relations of \cite{MR2165823} follow immediately from
our Equation \mref{eyb1}.

The defect/defect scattering matrix $U$ is defined in \cite{MR2165823} via an equation of the for $UTT=TTU$. In our picture $U$ corresponds to the intertwiner $\cR$ satisfying \mref{ddreln}. It may be read off from \mref{ersoln} in the special case $r_1=r_2=0$ corresponding to the type $I$ identification \mref{tident}. As mentioned in Section 2, the identification \mref{tident} yields two independent solutions for $U$ in this special case. Further physical requirements will be needed in order to pin down the defect/defect scattering matrix, but our method clearly provides a set of candidates from which such a scattering matrix may be selected.

\section{Discussion}

In this paper, we have developed the representation theory of generalised oscillator algebras in the $\uq$ case. We have produced  $\cL$ and $\cR$ operators, and demonstrated their use in the theory of lattice and sine-Gordon defects. 
We have made the explicit connections \mref{tident} and \mref{tiident} between our $\cL$ operator and the results
of CZ on type $I$ and type $I\!I$ 
quantum defects in the sine-Gordon model. 

There is a discussion in \cite{CZ10} of the appearance of the soliton/soliton S-matrix as a sub-block of the type $I\!I$ defect/soliton S-matrix for special defect parameter choices. In the algebraic picture of this current paper, this fact follows as  a  consequence of the finite truncation of $W_\z^{(\ur)}$ mentioned in Section 2.2.1.
Specifically, $W_\z^{(q,-q^{\pm 2},-q^{\pm 2})}$ truncates to a 
module which is $\uqbp$ isomorphic to the spin-$\half$ module $V_{\z q^{\pm 1}}$ associated with a soliton. 
The ability to draw such conclusions with minimal calculation, and the possibility of generalising the overall construction to other Toda theories are both reasons why a solid algebraic framework is useful in the analysis of defects.

\subsection*{Acknowledgements}
The author would like to thank Ed Corrigan and Cristina Zambon for bringing his attention to their paper \cite{CZ10} and for explaining their work. He would also like to thank Christian Korff for several
interesting conversations about defects, and to acknowledge the helpful comments of Zoltan Bajnok, Vincent Caudrelier and Anastasia Doikou on an earlier draft of this paper. He is also grateful to D. Binosi and L. Theu\ss l for writing and making available the Feynman graph plotting tool JaxoDraw.


\begin{thebibliography}{10}

\bibitem{JM}
M.~Jimbo and T.~Miwa.
\newblock {\em Algebraic Analysis of Solvable Lattice Models}.
\newblock CBMS Regional Conference Series in Mathematics, vol. 85. Amer. Math.
  Soc., 1994.

\bibitem{KMT05}
N.~Kitanine, J.~M. Maillet, N.~A. Slavnov, and V.~Terras.
\newblock {Correlation functions of the $XXZ$ spin-$\frac{1}{2}$ Heisenberg
  chain: recent advances.}
\newblock {\em Int. J. Mod. Phys.}, A 19:248--266, 2004.

\bibitem{MR1976005}
Joseph~A. Minahan and Konstantin Zarembo.
\newblock The {B}ethe-ansatz for {$ N=4$} super {Y}ang-{M}ills.
\newblock {\em J. High Energy Phys.}, (3):013, 29, 2003.

\bibitem{Caux08}
Jean-S\'{e}bastien Caux, Jorn Mossel, and Isaac~P\'{e}rez Castillo.
\newblock {The two-spinon transverse structure factor of the gapped Heisenberg
  antiferromagnetic chain}.
\newblock {\em JSTAT}, 0806.3069, 2008.

\bibitem{Bax72a}
R.J. Baxter.
\newblock {Partition Function of the Eight-Vertex Lattice Model}.
\newblock {\em {Annals of Physics}}, 70:193--228, 1972.

\bibitem{Bax73aI}
R.J. Baxter.
\newblock {Eight-Vertex Model in Lattice Statistics and One-Dimensional
  Anisotropic Heisenberg Chain. 1. Some Fundamental Eigenvectors}.
\newblock {\em Annals of Physics}, 76:1--24, 1973.

\bibitem{Bax73aII}
R.J. Baxter.
\newblock {Eight-Vertex Model in Lattice Statistics and One-Dimensional
  Anisotropic Heisenberg Chain. 11. Equivalence to a Generalized Ice-type
  Model}.
\newblock {\em Annals of Physics}, 76:25--47, 1973.

\bibitem{Bax73aIII}
R.J. Baxter.
\newblock {Eight-Vertex Model in Lattice Statistics and One-Dimensional
  Anisotropic Heisenberg Chain. 111. Eigenvectors of the Transfer Matrix and
  Hamiltonian}.
\newblock {\em Annals of Physics}, 76:48--71, 1973.

\bibitem{MR1384140}
Vladimir~V. Bazhanov, Sergei~L. Lukyanov, and Alexander~B. Zamolodchikov.
\newblock Integrable structure of conformal field theory, quantum {K}d{V}
  theory and thermodynamic {B}ethe ansatz.
\newblock {\em Comm. Math. Phys.}, 177(2):381--398, 1996.

\bibitem{BLZb}
V.V. Bazhanov, S.L. Lukyanov, and A.B. Zamolodchikov.
\newblock Integrable structure of conformal field theory {I}{I}: ${Q}$-operator
  and {D}{D}{V} equation.
\newblock {\em Comm. Math. Phys.}, 190:247--278, 1997.

\bibitem{Bazhanov:1998dq}
Vladimir~V. Bazhanov, Sergei~L. Lukyanov, and Alexander~B. Zamolodchikov.
\newblock {Integrable structure of conformal field theory. III: The Yang-Baxter
  relation}.
\newblock {\em Commun. Math. Phys.}, 200:297--324, 1999.

\bibitem{MR1832065}
Vladimir~V. Bazhanov, Sergei~L. Lukyanov, and Alexander~B. Zamolodchikov.
\newblock Spectral determinants for {S}chr\"odinger equation and {${\bf
  Q}$}-operators of conformal field theory.
\newblock In {\em Proceedings of the {B}axter {R}evolution in {M}athematical
  {P}hysics ({C}anberra, 2000)}, volume 102, pages 567--576, 2001.

\bibitem{Rossi:2002ed}
Marco Rossi and Robert Weston.
\newblock {A Generalized Q operator for $\uq$ vertex models}.
\newblock {\em J. Phys.}, A35:10015--10032, 2002.

\bibitem{MR2184024}
Christian Korff.
\newblock Solving {B}axter's {$TQ$}-equation via representation theory.
\newblock In {\em Noncommutative geometry and representation theory in
  mathematical physics}, volume 391 of {\em Contemp. Math.}, pages 199--211.
  Amer. Math. Soc., Providence, RI, 2005.

\bibitem{MR1880098}
Vladimir~V. Bazhanov, Anthony~N. Hibberd, and Sergey~M. Khoroshkin.
\newblock Integrable structure of {$W_3$} conformal field theory, quantum
  {B}oussinesq theory and boundary affine {T}oda theory.
\newblock {\em Nuclear Phys. B}, 622(3):475--547, 2002.

\bibitem{MR2426018}
Takeo Kojima.
\newblock Baxter's {$Q$}-operator for the {$W$}-algebra {$W_N$}.
\newblock {\em J. Phys. A}, 41(35):355206, 16, 2008.

\bibitem{MR757857}
A.~A. Belavin, A.~M. Polyakov, and A.~B. Zamolodchikov.
\newblock Infinite conformal symmetry in two-dimensional quantum field theory.
\newblock {\em Nuclear Phys. B}, 241(2):333--380, 1984.

\bibitem{MR858661}
John~L. Cardy.
\newblock Effect of boundary conditions on the operator content of
  two-dimensional conformally invariant theories.
\newblock {\em Nuclear Phys. B}, 275(2):200--218, 1986.

\bibitem{MR1279145}
G.~Delfino, G.~Mussardo, and P.~Simonetti.
\newblock Statistical models with a line of defect.
\newblock {\em Phys. Lett. B}, 328(1-2):123--129, 1994.

\bibitem{MR1674671}
Robert Konik and Andr{\'e} LeClair.
\newblock Purely transmitting defect field theories.
\newblock {\em Nuclear Phys. B}, 538(3):587--611, 1999.

\bibitem{MR2087101}
P.~Bowcock, E.~Corrigan, and C.~Zambon.
\newblock Classically integrable field theories with defects.
\newblock In {\em Proceedings of 6th {I}nternational {W}orkshop on {C}onformal
  {F}ield {T}heory and {I}ntegrable {M}odels}, volume~19, pages 82--91, 2004.

\bibitem{MR2045593}
Peter Bowcock, Edward Corrigan, and Cristina Zambon.
\newblock {Affine {T}oda field theories with defects}.
\newblock {\em J. High Energy Phys.}, 1:056, 25 pp. (electronic), 2004.

\bibitem{MR2165823}
Peter Bowcock, Edward Corrigan, and Cristina Zambon.
\newblock {Some aspects of jump-defects in the quantum sine-{G}ordon model}.
\newblock {\em J. High Energy Phys.}, 8:023, 35 pp. (electronic), 2005.

\bibitem{MR2326786}
Edward Corrigan and Cristina Zambon.
\newblock {On purely transmitting defects in affine {T}oda field theory}.
\newblock {\em J. High Energy Phys.}, 7:001, 38 pp. (electronic), 2007.

\bibitem{CoZa09}
Edward Corrigan and Cristina Zambon.
\newblock {A New Class of Integrable Defects}.
\newblock {\em J. Phys.}, A42:475203, 2009.

\bibitem{CZ10}
E.~Corrigan and C.~Zambon.
\newblock {A transmission matrix for a fused pair of integrable defects in the
  sine-Gordon model}.
\newblock arXiv:1006.0939, 2010.

\bibitem{MR2398297}
Ismagil Habibullin and Anjan Kundu.
\newblock Quantum and classical integrable sine-{G}ordon model with defect.
\newblock {\em Nuclear Phys. B}, 795(3):549--568, 2008.

\bibitem{MR2469290}
V.~Caudrelier.
\newblock On a systematic approach to defects in classical integrable field
  theories.
\newblock {\em Int. J. Geom. Methods Mod. Phys.}, 5(7):1085--1108, 2008.

\bibitem{Baj06}
Z.~Bajnok.
\newblock {Equivalences between spin models induced by defects }.
\newblock {\em JSTAT}, page p06010, 2006.

\bibitem{Baj08}
Z.~Bajnok and Zs. Simon.
\newblock {Solving topological defects via fusion }.
\newblock {\em Nucl. Phys.B}, 802:307--329, 2008.

\bibitem{ChPr91}
V.~Chari and A.~Pressley.
\newblock {Quantum Affine Algebras}.
\newblock {\em Comm. Math. Phys.}, 142:261--283, 1991.

\bibitem{Ku91}
P.P. Kulish.
\newblock {Contraction of quantum algebras, and $q$-oscillators}.
\newblock {\em Theor. and Math. Phys}, 86:108--110, 1991.

\bibitem{MR2472021}
H.~Boos, M.~Jimbo, T.~Miwa, F.~Smirnov, and Y.~Takeyama.
\newblock Hidden {G}rassmann structure in the {$XXZ$} model. {II}. {C}reation
  operators.
\newblock {\em Comm. Math. Phys.}, 286(3):875--932, 2009.

\bibitem{BJMST07}
H.~Boos, M.~Jimbo, T.~Miwa, F.~Smirnov, and Y.~Takeyama.
\newblock {{Fermionic} Basis for the Space of Operators in the XXZ Model}.
\newblock {\em SISSA Proceedings of Science}, 1:015, 34 pp. (electronic), 2007.

\bibitem{MR1688868}
Anjan Kundu.
\newblock Algebraic approach in unifying quantum integrable models.
\newblock {\em Phys. Rev. Lett.}, 82(20):3936--3939, 1999.

\end{thebibliography}

\end{document}